# VISUAL SPATIAL LEARNING OF COMPLEX OBJECT MORPHOLOGIES THROUGH INTERACTION WITH VIRTUAL AND REAL-WORLD DATA

Chiara Silvestri#Δ, René MotroΔ, Bernard MaurinΔ and

Birgitta Dresp-LangleyΔ*

*ΔEquipe SLA, Ecole Nationale Supérieure d'Architecture de Montpellier – France*
*#Dipartimento Costruzione dell'Architettura, I.U.A.V., Venezia – Italy*
*\*UMR 5508 CNRS, Université de Montpellier, Montpellier – France*

\*To whom correspondence should be addressed. E-mail: dresp@lmgc.univ-montp2.fr

## ABSTRACT

To shed light on the visual perceptual processes which may explain why virtual reality data can be more effective than real-world data when dealing with complex, steadily changing object worlds, we conducted a visual-spatial memory experiment with 48 volunteers. 24 of them were experts in conceptual spatial design and 24 were non-expert observers. None were particularly familiar with the abstract visual structure presented as three different models: 2-D single view, 3-D virtual with random multiple views, 3-D real with random multiple views). When asked to draw elements of the visual structure from memory, expert designers were found to perform significantly better after having explored multiple views of the virtual structure, compared with other experts who explored multiple views of the real-world model of the same structure. Comparing performances between the two study populations, we found expertise to produce a significant advantage of the two 3-D-random-multiple-view conditions, particularly in the condition of virtual viewing. It is concluded that interacting with virtual reality data facilitates the perceptual processing of complex visual structures in design experts highly familiar with virtual 3D rendering software. We suggest that this is made possible through specific eligibility traces made available by interaction in virtual 3D for complex many-to-one memory matching operations.

# INTRODUCTION

In the digital age, the role of visual learning and communication is becoming more and more important. Computer driven image technology and virtual reality devices have replaced former written or spoken media of inter-personal information exchange in a wide range of public and private domains, such as education, healthcare, and navigation (e.g. Katz *et al.*, 2006, Stelzer & Wickens, 2006). In the domains of architecture and building design, current ecological pressure and a rapidly changing global society have placed a premium on new ways of conceiving objects in space, with a general motivation for the most rational and sustainable, but also imaginative, expressive and beautiful solutions. The conceptual design of novel and complex object structures in the real world relies on the expert's learnt ability to effectively manipulate mental representations of visual three-dimensional (3-D) space. In design learning as in many other fields, virtual data have largely replaced former media of real-world learning (e.g. Borgart & Kocaturk, 2007). This shift towards new forms of communication between humans and virtual environments is bound to have some, not yet fully measurable, impact on the development of individuals and society. Only little is still known about the visual-perceptual processes through which virtual reality data may guide information processing in comparison with real-world data. Studies on learning programs and skill acquisition in surgical training, for example, have shown that the learning and practice of minimally invasive surgery through virtual reality imaging tools significantly improves surgical technical skills, leading to a more rapid skill transfer and generalization compared with real-world training methods (e.g. Gallagher *et al.*, 2005). Virtual reality is often referred to in terms of 'augmented reality' producing enriched perceptual environments that are free from the information constraints of real-world data. Experimental investigations into cognitive processes of learning and memory, for example, have shown that virtual reality provides a learning medium with greater efficiency than a real-world situation, which is limited by the constraint of partially observable data and solicits more heavily the experience and prior knowledge of the learning individual (e.g. Matheis *et al.*, 2007).

A certain number of visual perceptual studies have examined the precision and speed with which human observers' recognize local aspects of three-dimensional shape structures, usually with familiar or so-called natural objects (see Norman *et al*, 2004, for a review). Our study here is focussed on less well studied aspects of perceptual representation by addressing questions relative to the functional role of specific geometric properties of novel object structures in perceptual learning.

*Perceptual learning by agent-environment interaction*

The perceptual learning of abstract, novel structures may be approached through concepts of machine learning and skill acquisition by agents, such as the concept of representation matching (e.g. Carpenter & Grossberg, 1991, Whitehead & Ballard, 1991) and the concept of partially observable worlds (e.g. Singh & Sutton, 1996). It is stipulated that any agent, man or machine, is capable of learning from its environment, be it virtual or real, on the basis of perceptual and entirely non-verbal actions (see Figure 1 below).

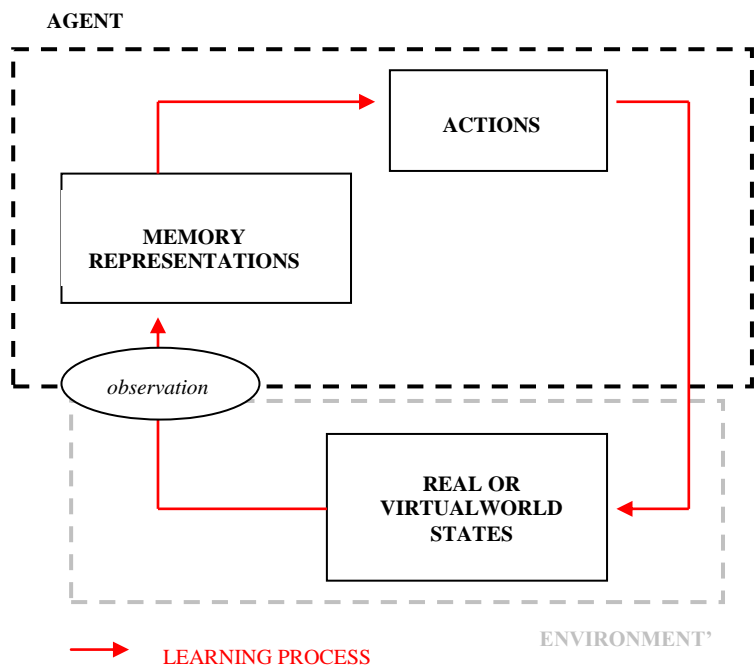

Figure 1: An agent may communicate with an environment, be it virtual or real, through visually guided mental operations where observations are matched to memory representations of world states. This learning process enables action upon steadily changing, real or virtual, world states.

A perceptual action describes a formal or mental operation where a given or several states of the real or virtual world is/are matched to observations of the latter. A matching operation usually triggers an action, like moving the cursor to a new position on the computer screen in a virtual reality situation, or pulling a handle to open a drawer of a filing cabinet in a real-world scenario. The criteria for matching already learnt states of the world to new observations, the so-called learning criteria, exploit what researchers in the field of machine learning refer to as *eligibility traces* (Singh & Sutton, 1996). Eligibility traces are working memory data, with a specific heuristic or diagnostic value, which correspond to representations of simple or complex events and actions. In the context of the present study, two types of matching operations will be considered: 'one-to-one' matching and 'many-to-one' matching.

*One-to-one matching*

In one-to-one matching, a unique observation is matched to a unique world-state. In the real-world context, this could correspond to the situation where an agent matches the front of a building pointed out to him/her in a brochure or leaflet to that of a building visited earlier when travelling. In a virtual reality context, one-to one matching could correspond to a situation where an agent matches a corridor of a virtual building to the only corridor known from previous experience to lead to the exit. Such a match may then be followed by entering that specific corridor. In the case of one-to-one category matching, a single eligibility trace, like the colour of the corridor represented on the screen, or several eligibility traces (colour + width + length) may be used.

*Many-to-one matching*

In many-to-one matching, several observations are matched to a unique world-state. In the real world, the case of many-to-one matching could correspond to a situation where an art expert identifies several sculptures in an auction house as representing the work of a single, particular artist. The same example can be directly translated into a virtual reality situation, where the expert would make the same match while navigating through a virtual gallery where these art objects are represented. The eligibility traces for a many-to-one match may be multiple (artist's preferred theme + combination of colours used + medium used + period estimate) or single, such as in the case of a match on the sole basis of the artist's preferred theme, for example. Successfully matching many observations to a single world state requires knowledge (expertise) of specific eligibility traces for a potential match. We assume that matching operations as described here constitute the basis of the perceptual learning process through which knowledge about the structure of complex visual objects is made accessible. An example of specific visual perceptual eligibility traces for learning complex object structure can be given on the basis of Euclidian geometrical principles describing spatial relations between elements of a 'simplex' structure.

*Eligibility traces in the simplex structure*

The simplex is an abstract design object which has a characteristic three-fold symmetrical structure and consists of three rigid bars held together by tensed cables. The few structural design engineers and architects who have actually worked with it, consider the simplex, as may be guessed from the name, the simplest complex structure hitherto invented (e.g. Snelson, 1965, Motro, 2003). Seven geometrical descriptors account for the ways in which the elements of this structure are related to each other in the plane:

- three lines connect three oblique bars one-to-one at their two ends
- these lines form the shape of a triangular surface at each end
- one such triangle is the planar projection by translation of the three end-points of the other triangle, with a 30° rotation along the vertical/lateral axis of symmetry of the structure
- three lines with a 30° tilt connect the triangles one by one at their end-points
- any two such lines together with the two sides of the triangles they connect form the shape of one of three polygonal surfaces
- the vertical/lateral axis of symmetry of the structure is a virtual line connecting the central points of the triangular surfaces, or the central points of two polygonal surfaces, depending on how the structure is oriented in the plane
- the three oblique bars are arranged symmetrically around these virtual lines (three-fold symmetry)

The conditions under which the simplex is viewed and/or explored by observers who have never seen it may determine whether or not these Euclidian properties, or eligibility traces, are effectively made accessible to the perceptual learning process underlying the formation of a structural representation.

*Virtual viewing and exploration by vision and touch*

Structural symmetry is an important factor in the visual-spatial processing of virtually viewed objects. A single virtual sample view may facilitate the recognition of a visual object presented in a novel view when the object is bilaterally symmetrical. It is assumed that human observers may, in this case, be able to derive representations of additional views from a single virtual sample view on the basis of symmetry transformations (e.g. Vetter, Poggio, & Bülthoff, 1994). In the case of the simplex, this could mean that the threefold symmetry of the central bars of the structure facilitates access to representations of the triangular and polygonal surfaces connecting their ends.

Another important factor in visual spatial processing of structure could be the sensory modality through which a structure is explored by a human observer. Earlier views, such as Gibson's (1962, 1963, 1966), considered visual and tactile exploration as equivalent media, making essentially the same kind of information available to an observer. As a consequence, visual-plus-tactile exploration of a novel object would not provide more information compared with purely visual exploration. Other more recent studies suggest that this may not necessarily be the case, like in shape recognition experiments where observers were found to recognize target shapes presented among other shapes significantly better when the target was previously explored visually and by hand rather than having been viewed only (Norman *et al*, 2004). Thus, structural information processing may well be facilitated when observers are able to not only see but also touch and explore an object by hand.

In the case of the simplex, visual exploration of all possible 2-D views of the structure should in principle suffice to give access to planar representations of the geometrical eligibility traces listed above (Euclidian or planar eligibility traces). On the other hand, visual-plus-tactile exploration of the real 3-D object might facilitate access to these eligibility traces by providing additional cues to structure that are not made available through mere viewing. When exploring the simplex by hand, observers become aware of some of its mechanical properties, such as the tension in the cables defining the triangular and polygonal surfaces, and the link between this tension and the 30° rotation of the triangles connecting the bars of the structure at their ends. Thus, when given the opportunity to explore the simplex manually, observers might gain an advantage for understanding its structure compared with observers who are only given the chance to explore multiple 2-D views.

A viewpoint-dependence of spatial information processing in object recognition has been found with familiar objects, where visual recognition was best when objects were viewed from the front, and tactile recognition was best when the back of objects was explored manually (Newell, Ernst, Tjan, & Bülthoff, 2001). The axis of rotation appears to be a critical factor determining viewpoint dependence in the visual modality. Visual viewpoint dependence is abolished by perceptual learning, or through repeated interaction with virtual 3-D environments (e.g. Christou & Bulthoff, 1999). Visual-tactile recognition is viewpoint-independent, even for unfamiliar objects (Lacey, Peters, & Sathian, 2007), which indicates that object structures accessed by multimodal recognition are formed at higher cognitive levels of information processing and transformation.

Drawing a visual object from memory fully solicits such higher cognitive levels. Recent research has shown that the act of drawing is a powerful means of accessing, activating and consolidating knowledge representations of object properties in the memory structures of the right brain hemisphere, involving the most important functional regions for learning and communication, such as the Brodman area (Harrington *et al.*, 2006). To successfully draw a novel object or parts of it from memory involves cognitive processes of attention and capacity-limited working memory.

*Attention and working memory*

The well-defined capacity limits of selective processes of attention and working memory have been established on the basis of various studies (Oberly, 1928, Miller, 1956, Parkin, 1999, Vogel, Woodman & Luck, 2001), which have all shown that any normally developed human adult is capable of attending to an average maximum of seven (7 +/- 2) items or representations, and to maintain these representations within working memory for up to several hours if necessary (cf. Potter, 1993). Representations maintained in working memory may correspond to single numbers, words, meaningful concepts, visual scenes, or fragments such as parts of a specific object structure. Attending to and maintaining within working memory either its seven elementary parts (two triangles, three polygons, and three central bars), or the seven geometrical principles (listed above) defining its structural composition should not overtax the attention span and working memory capacity of normally developed adults.

*Effects of ethnicity and gender*

Recent findings collected from an isolated Amazonian indigene group suggest that elementary geometrical principles are part of the universal core knowledge (e.g. Dehaene, Izard, Pica & Spelke, 2006) that is present in all humans, regardless of education or gender. Although some visual-spatial information processing tasks, such as mental rotation tasks, sometimes show a male performance advantage, other recent studies have shown that such gender differences do not extend to visual-spatial processing in general, and may largely depend on the specific conditions of a specific test (e.g. Seurinck, Vingerhoets, de Lange, & Achten, 2004).

**METHODS**

*Observers*

The 24 non-experts consisted of 16 young male and 8 young female adults who were students in biology or human sciences. In a short interview before the experiment, we made sure that none of them was an experienced user of 3-D image processing software tools or regularly played computer or video games. The 24 expert subjects consisted of 22 young male and 2 young female design engineers and/or architects who were all experienced users of computer tools for 3-D shape representation.

*Materials*

Three different model representations of the 'simplex' (Figure 2) were generated for the three different test conditions of the experiment. The first model consisted of a single 2-D view of the structure, with three coloured bars (red, yellow, blue) connected by dark cables in one case, and three black bars connected by dark cables in the other. These 2-D model single-view images were printed on sheets of paper and presented to subjects in the test condition that is referred to here as "2-D single-view".

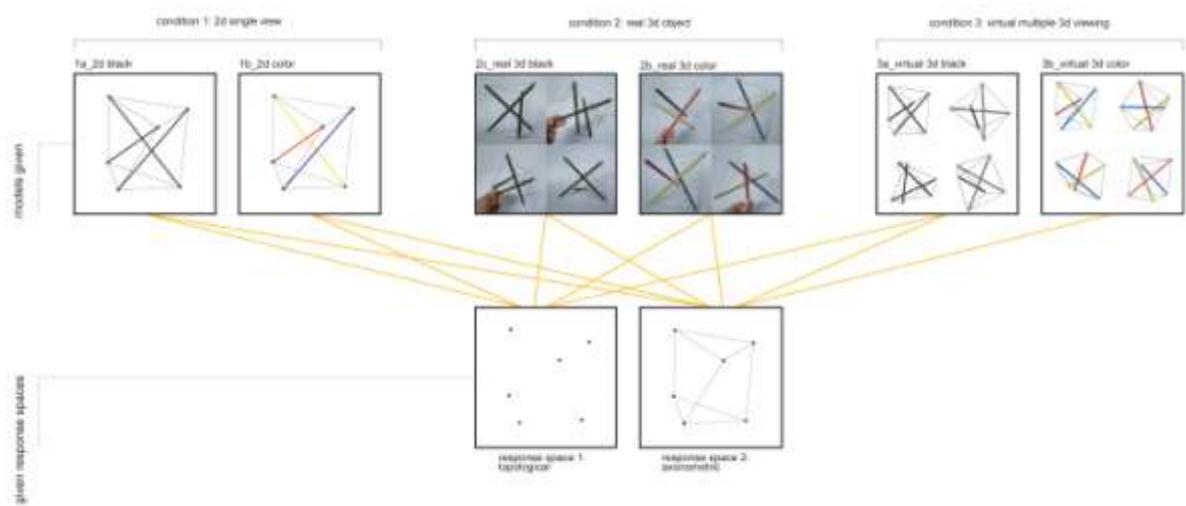

Figure 2: Different model and viewing conditions and reference frames given on the response sheets.

The second model consisted of a real '*tensegrity simplex*' made of three wooden bars, held together by black nylon cables. As in the other two models, the three bars were coloured in one case and black in the other. The subjects could hold and actively manipulate the model for as long as they wanted in this condition (see videos 1 and 2 associated with Figure 2), referred to here as "real 3-D active perception".

The third model consisted of a "multiple-view" version of the same 2-D model. It was presented to subjects in the test condition referred to as "virtual 3-D multiple viewing". In this condition, the observers were placed in front of a randomly chosen view of the structure on a computer screen and given the opportunity to generate multiple views of the structure by using the mouse button (see videos 3 and 4 associated with Figure 2). The "2-D single-view" and "virtual 3-D multiple-views" models were generated through 'AUTOCAD Architectural Desktop 2006'.

*Procedure*

The 24 experts and the 24 non-experts were divided into three groups of eight observers for each population. Each group of subjects was given only one of the three models described above. To eliminate serial effects, half of the observers of each group were shown a version of the model with coloured bars in the first round and the version with black bars in the second round, the other half were given the model with black bars first. Subjects had the opportunity to inspect a given model for as long as they wanted and were informed that they would be required to draw certain parts of the model structure from memory. When they felt ready, the model was taken away and they were asked to draw the three bars of the structure on the two separate response sheets with the two different reference frames. Again, to eliminate possible sequence effects, the presentation order of the response sheets was counterbalanced between observers of a given group. Also, the order in which the response sheets were handed to a given observer was counterbalanced between successive tests (i.e. between "coloured bars first round" and "black bars second round" and between "black bars first round" and "coloured bars second round"). The times taken by an observer to draw the three bars, measured with a chronometer, and the number of errors made were recorded. Bars drawn at the appropriate position within the reference frame were counted as "correct" and bars drawn at inappropriate positions were counted as "errors".

Before testing the two study populations, we investigated the performances of three 'super-experts' who were particularly familiar with the conceptual design of tensegrity structures and therefore excluded from the experiment. To get an idea of the optimal performance level in the task proposed here, these 'super-experts' were tested in the three experimental conditions (one super-expert for each 'model' condition). They all drew the requested visual elements from memory in times between 5 and 20 seconds with no errors, regardless of the type of model shown, the colour of the model, or the spatial reference frame provided for drawing. These results were not included in the data presented here.

**RESULTS**

Average times (measured in seconds) taken to draw the three bars from memory, total and average numbers of positional errors made, and the corresponding standard deviations were computed for each of the two populations studied and the different test conditions. ANOVA was performed to assess the statistical significance of differences between means observed. In a first analysis, 'between-groups' ANOVA relative to times and errors was performed to assess the global effects of study population and models given to observers. This includes analyses of the interaction of each of the two main factors with the spatial reference frame provided for drawing ('topological' versus 'axonometric') and the colour of the bars of the model structure ('three different colours' versus 'black only'), as well as of the interaction between these two additional factors.

*Global effects of study population and type of model environment*

The data show, as expected, that the average time for reproducing the three bars of the visual 3-D structure by drawing from memory was significantly shorter for conceptual designers (F (1,190) = 10.4850, p < .01). Also, the number of topological errors, with a total of three possible topological errors per drawing, was significantly smaller for the experts (F (1,190) = 31.0960, p < .001). Average times taken to draw from memory were the shortest and errors made the least when the '2-D single-view model' was shown to observers. Times were the longest with the 'real 3-D active perception model', and the most errors were made, globally, with the 'virtual 3-D multiple-view' model. These global effects of the type of model given to observers before drawing from memory was statistically significant on both times taken to draw (F (2,189) = 16.5660, p < .001) and on the number of errors made (F (2,189) = 9.2440, p < .001). Post-hoc analyses of the effect of the three models on the times taken to draw revealed all comparisons between all levels of this factor to be statistically significant (*t* (1,126) = 5.7260, p < .001 for "real 3-D active perception" *vs* "2-D single-view"; *t* (1,126) = 2.3510, p < .05 for "real 3-D active perception" *vs* "virtual 3-D multiple-views"; *t* (1,126) = 3.3750, p < .001 for "2-D single-view" *vs* "virtual 3-D multiple-views"). Post-hoc analyses of the effect on errors revealed two of the comparisons as statistically significant (*t* (1,126) = 3.5170, p < .001 for "real 3-D active perception" *vs* "2-D single-view", $t(1,126) = 3.9010$, $p < .001$ for "virtual 3-D multiple-views" *vs* "2-D single-view", $t(1, 126) = 0.3840$, $p = .70$ NS for "real 3-D active perception" *vs* "virtual 3-D multiple-views"). The interaction between the 'population factor' and the 'model factor' was statistically significant for times ($F(5,186) = 13.2420$, $p < .001$) and for errors ($F(5,186) = 14.5840$, $p < .001$). Average times and numbers of errors per condition with their standard deviations ($\sigma$) are summarized below.

|  | **Experts** | | **Non-experts** | |
|---|---|---|---|---|
|  | Errors | Times | Errors | Times |
| '2-D single-view' | m = 0.03<br>$\sigma$ = 0.17 | m = 23<br>$\sigma$ = 14.7 | m = 0.28<br>$\sigma$ = 0.63 | m = 14<br>$\sigma$ = 6.9 |
| 'Virtual 3-D multiple-view' | m = 0.12<br>$\sigma$ = 0.49 | m = 22<br>$\sigma$ = 12.6 | m = 1.41<br>$\sigma$ = 1.13 | m = 60<br>$\sigma$ = 53.3 |
| 'Real 3-D active perception' | m = 0.43<br>$\sigma$ = 0.80 | m = 43<br>$\sigma$ = 37.3 | m = 0.97<br>$\sigma$ = 1.09 | m = 69<br>$\sigma$ = 51.4 |

Average times taken to draw and total numbers of errors (not average values as given above) are represented as histograms below (Figures 3a and b, respectively).

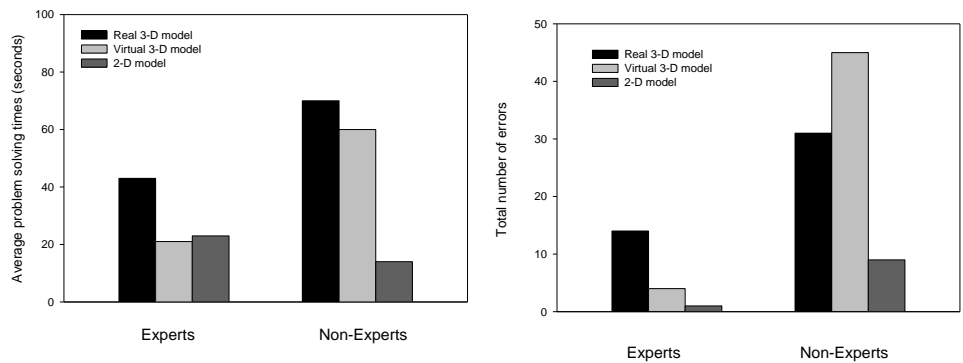

Figure 3: Average times taken to draw from memory, here referred to as problem solving times (left), and the total number of errors made (right) plotted as a function of the study population and the three different model environments.

*Interactions with spatial reference frame and colour*

Observers drew generally faster and made fewer errors when axonometric rather than merely topological information was provided on the response sheet. When the bars of the model structure were all black instead of having different colours, drawing the bars from memory was generally more rapid and fewer errors were made. While the global effect of the spatial reference frame for drawing (topological *vs* axonometric), the colour of the bars of the model structures (three colours *vs* all black), and their interaction did not produce statistically significant differences, significant interactions with the 'population factor' and with the 'model factor' were found.

The facilitating effect of the axonometric reference frame, compared with the merely topological one, was considerably stronger in the non-expert population, in regard to both the average time taken to draw from memory and the average number of errors made. While such facilitating effects were also present in the expert population, they were noticeably weaker. This interaction between the 'population factor' and the spatial reference frame provided for drawing was statistically significant, for both times ($F(3,188) = 4.3210$, $p < .01$) and for errors ($F(3, 188) = 10.7570$, $p < .001$). Average times and numbers of errors per condition for this interaction with the corresponding standard deviations ($\sigma$) are summarized below.

|  | **Experts** | | **Non-experts** | |
| --- | --- | --- | --- | --- |
|  | Errors | Times | Errors | Times |
| 'axonometric reference frame' | m = 0.12<br>$\sigma$ = 0.44 | m = 28<br>$\sigma$ = 24.5 | m = 0.79<br>$\sigma$ = 1.05 | m = 41<br>$\sigma$ = 33.3 |
| 'topological reference frame' | m = 0.27<br>$\sigma$ = 0.67 | m = 30<br>$\sigma$ = 27.7 | m = 0.97<br>$\sigma$ = 1.10 | m = 54<br>$\sigma$ = 60.2 |

Similarly, the facilitating effects of uniformly black bars of a model structure, compared with coloured ones, were generally more pronounced in the non-expert population. This interaction between the 'population factor' and the colour of the bars of the models was significant for times ($F(3, 188) = 3.7140$, $p < .05$) as well as for errors ($F(3, 188) = 11.0500$, $p < .001$). Average times and numbers of errors per condition for this interaction with the corresponding standard deviations ($\sigma$) are summarized below.

|  | **Experts** | | **Non-experts** | |
|---|---|---|---|---|
|  | Errors | Times | Errors | Times |
| 'models with black bars' | m = 0.19<br>σ = 0.57 | m = 27<br>σ = 26.5 | m = 0.77<br>σ = 1.05 | m = 43<br>σ = 39.8 |
| 'models with coloured bars' | m = 0.21<br>σ = 0.58 | m = 31<br>σ = 25.8 | m = 1.01<br>σ = 1.09 | m = 49<br>σ = 56.8 |

When comparing the facilitating effects of the axonometric reference frame of the different models, we observed the strongest facilitation, both in terms of average times taken to draw and errors made, for the 'real 3-D active perception' model. The same facilitation, but to a lesser extent, was found with the 'virtual 3-D multiple-view' model. With the '2-D single-view' model, no facilitation of the axonometric reference frame compared with the topological one, on either times or errors made, was found. This interaction between the 'model factor' and the spatial reference frame provided for drawing was statistically significant for times (F (5, 186) = 7.6390, p < .001) and for errors (F (5, 186) = 4.2620, p < .01). Means and standard errors are summarized below.

|  | **2-D single-view** | | **Real 3-D active** | | **Virtual 3-D multi-view** | |
|---|---|---|---|---|---|---|
|  | Errors | Times | Errors | Times | Errors | Times |
| 'axonometric' | m = 0.15<br>σ = 0.45 | m = 18<br>σ = 11.1 | m = 0.53<br>σ = 0.87 | m = 46<br>σ = 36.1 | m = 0.69<br>σ = 1.09 | m = 40<br>σ = 29.7 |
| 'topological' | m = 0.16<br>σ = 0.51 | m = 18<br>σ = 13.4 | m = 0.88<br>σ = 1.07 | m = 66<br>σ = 53.5 | m = 0.84<br>σ = 1.08 | m = 41<br>σ = 53.6 |

When examining the facilitating effects of uniformly black bars compared with coloured bars of the different models, summarized below, we observe the strongest facilitation on times with the '3-D active perception' model and the strongest effect on errors with the '3-D virtual multiple views' model. No facilitation of the black bars, on either times or errors, was found with the '2-D single-view' model. The interaction between the 'model factor' and the colour of the bars of the models was significant for times (F (5, 186) = 8.4080, p < .001) and for errors (F (5, 186) = 3.9880, p < .01).

|  | 2-D single-view | | Real 3-D active | | Virtual 3-D multi-view | |
|---|---|---|---|---|---|---|
|  | Errors | Times | Errors | Times | Errors | Times |
| 'black bars' | m = 0.15<br>σ = 0.51 | m = 18<br>σ = 10.5 | m = 0.62<br>σ = 0.97 | m = 58<br>σ = 45.9 | m = 0.65<br>σ = 1.03 | m = 39<br>σ = 26.9 |
| 'coloured bars' | m = 0.15<br>σ = 0.44 | m = 19<br>σ = 13.8 | m = 0.78<br>σ = 1.01 | m = 62<br>σ = 42.1 | m = 0.88<br>σ = 1.13 | m = 42<br>σ = 54.9 |

In the next step of the analysis, 'within-group' ANOVA, for times and errors was performed on the data of the two study populations.

*Expert conceptual designers*

Within the expert population, observers drew most rapidly from memory with the 'virtual 3-D multiple-views' model and, as in the '2-D single-view' condition, they made were very few errors. In contrast, the model for which drawing from memory took the longest and the most errors were made was the 'real 3-D active perception' model. The effect of the 'model factor' was significant within the expert population on both times ($F_{(2,93)} = 8.2010$, $p < .001$) and errors ($F_{(2,93)} = 4.6500$, $p < .05$). Post-hoc analyses revealed significant effects for 'real 3-D active perception' *vs* '2-D single-view' for times ($t_{(1,62)} = 3.3950$, $p < .01$) and errors ($t_{(1,62)} = 2.9070$, $p = .01$), and for 'real 3-D active perception' *vs* 'virtual 3-D multiple-views' ($t_{(1,62)} = 3.6100$, $p < .001$ and $t_{(1,62)} = 2.2530$, $p < .05$ for times and errors respectively). The comparison '2-D single-view' *vs* 'virtual 3-D multiple-views' was neither significant for times, nor for errors within the expert population, as could be expected from the small differences between means (see data on the left of Figures 3a and b, given above).

Experts were faster and made fewer errors when the axonometric reference frame was provided for drawing compared with the merely topological one. Also, they drew from memory faster and made fewer errors when the bars of the models shown were all black instead of coloured. While the facilitating effects of axonometric reference frames and monochrome structures, or interaction between them, were as such not statistically significant, significant interactions of either variable with the 'model factor' were found.

The facilitating effect of an axonometric reference frame compared with a topological one was the strongest, on both times and errors, in the 'real 3-D active' condition. Such facilitation was considerably weaker or absent in the two other conditions of the 'model factor'. This interaction between the spatial reference frame provided for drawing and the 'model factor' was statistically significant for times (F (5,90) = 3.2710, p < .01) and for errors (F (5,90) = 2.5890, p < .05). These data are shown in Figure 4.

The facilitating effect of black model bars compared with coloured ones on average times taken to draw was the strongest in the 'real 3-D active perception' and the 'virtual 3-D multiple views' conditions; this effect was considerably weaker in the '2-D single-view' condition. For errors, a slight facilitation of black model bars was observed with the 'real 3-D active' model. The interaction between the reference frame provided for drawing and the 'model factor' was statistically significant for times (F (5,90) = 3.4160, p < .01), but not for errors (F (5,90) = 1.9740, p = .09 NS), as expected from the small differences between means. These data are represented in Figure 5.

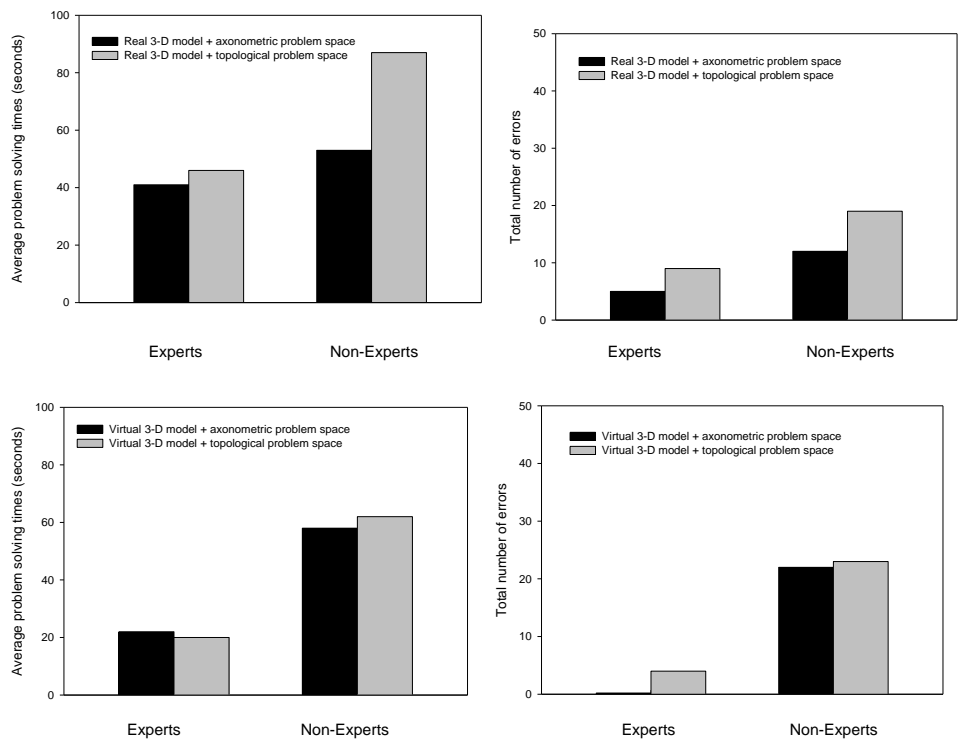

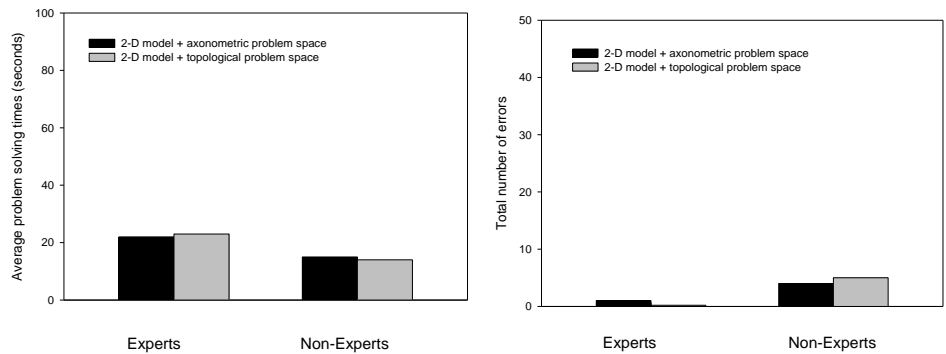

Figure 4: Average times taken to draw from memory (left) and the total number of errors made (right) plotted as a function of the study population and the spatial reference frame given for the different model conditions.

*Non-experts*

Within the non-expert population, observers drew most rapidly from memory and made the fewest errors in the '2-D single-view' condition. In contrast with the expert conceptual designers, the non-experts took considerably longer to draw from memory, and made the most errors in the 'virtual 3-D multiple-views' condition, where the experts were the fastest and most accurate compared with the other two model conditions. The effect of the 'model factor' was significant within the non-expert population for both times ($F(2,93) = 15.6330$, $p < .001$) and errors ($F(2,93) = 10.6800$, $p < .001$). Post-hoc analyses revealed significant effects of the comparison 'real 3-D active perception' *vs* '2-D single-view' for times ($t(1,62) = 5.2490$, $p < .001$) and errors ($t(1,62) = 2.7850$, $p = .01$), and for the comparison '2-D single-view' *vs* 'virtual 3-D multiple-views' ($t(1,62) = 4.2950$, $p < .001$ and $t(1,62) = 4.5870$, $p < .001$ for times and errors respectively). The comparison '3-D multiple-views' *vs* 'virtual 3-D multiple-views' was neither significant for times, nor for errors within the non-expert population, in contrast to what has been observed with the experts (see data on the right in Figures 3a and b, given above).

Non-expert observers were noticeably faster and made noticeably less errors when the axonometric reference frame was given rather than the merely topological one. Also, they drew from memory faster and made fewer errors when the bars of the models were all black instead of coloured. While these facilitating effects, and interactions between them, were as such not statistically significant, significant interactions with the 'model factor' were found. The facilitating effect of an axonometric reference frame compared with a topological one was the strongest, on both times and errors, in the 'real 3-D active perception' condition. This facilitation was slightly weaker in the 'virtual 3-D multiple-views' condition and noticeably weaker in the '2-D single-view' condition. The interaction between the spatial reference frame provided for drawing and the 'model factor' was statistically significant for times ($F(5,90) = 7.3700$, $p < .001$) and for errors ($F(5,90) = 4.5500$, $p < .001$). The data relative to these effects are shown in Figure 4.

The facilitating effect of black model bars compared with coloured ones on average times taken to draw was the strongest in the 'real 3-D active perception' and the 'virtual 3-D multiple views' conditions. The effect was absent in the '2-D single-view' condition. For errors, the strongest facilitation of black model bars was observed with the 'virtual 3-D multiple-views' model, followed by the 'real 3-D active' condition. A weak effect in the same direction was noted in the 2-D single-view condition. The interaction between the reference frame and the 'model factor' was statistically significant for times ($F(5,90) = 6.4150$, $p < .001$) and for errors ($F(5,90) = 4.6430$, $p < .001$), in contrast with the observations reported for the experts. These data are represented in Figure 5.

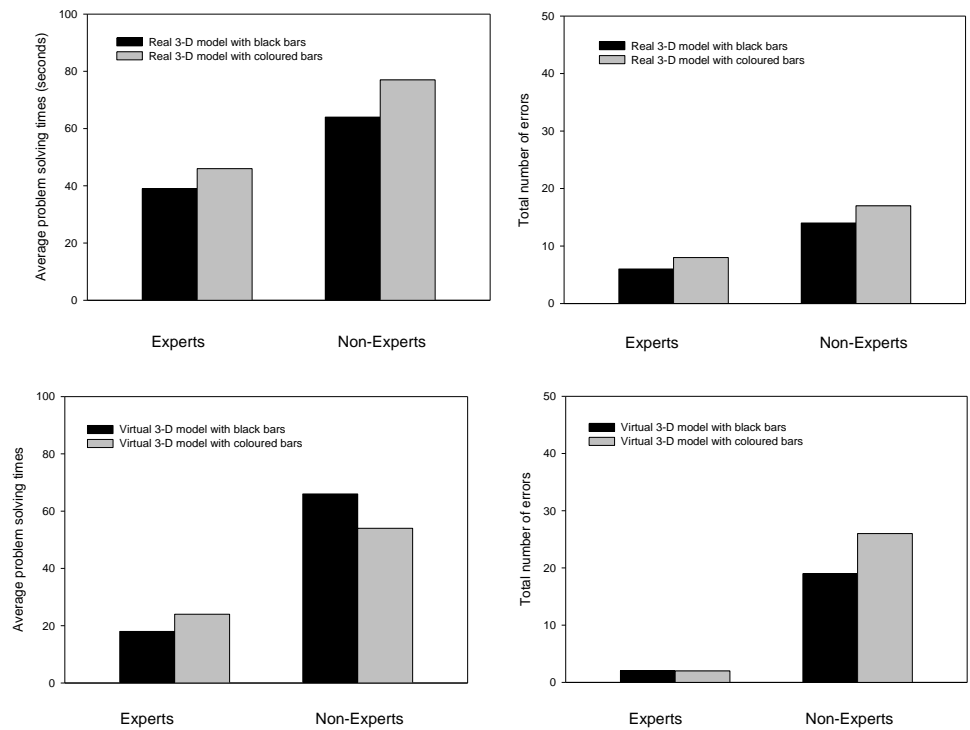

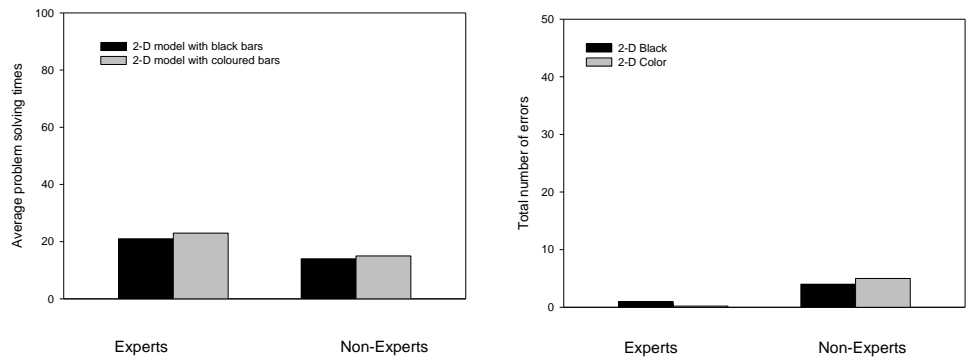

Figure 5: Average times taken to draw from memory (left) and the total number of errors made (right) plotted as a function of the study population and the model colour for the different model conditions

**DISCUSSION**

When asked to draw elements of an abstract, non-familiar visual structure from memory, expert designers perform significantly better after having explored multiple views of a virtual representation of the structure, compared with other experts who explored multiple views of a real-world model of the same structure. Also, their specific expertise or practice in the manipulation of complex visual structures generates a significant advantage, compared with non-experts, for visual-spatial memory matching subsequent to 3-D-multiple-viewing, particularly in the case of virtual viewing. Local visual information of colour seemed to perturb rather than facilitate this process given that performances were generally better in the monochrome viewing conditions, for both virtual and real-world models.

These observations shed light on the nature of the specific perceptual learning process that explains how humans learn to master abstract visual structural complexity through interaction with virtual reality data. Without any particular familiarity with the complex visual structure given, experts were perfectly able to maintain multiple virtual views of it in working memory and to match these short-term memory representations successfully to a single and totally unexpected view of the structure (see Figure 6). Non-experts could not perform this many-to-one matching, as revealed by their poor performances in the multiple viewing conditions, and in particular the virtual viewing condition.

Although it has been shown previously that humans and pigeons can acquire view-point-independent representations of complex visual objects through object-specific visual learning (Christou & Bulthoff, 1999, Christou *et al.*, 2003, Peissig et al., 2002,

Spetch & Friedman, 2003), it is unlikely that this explains why the experts, who were not more familiar with the visual structure than the non-experts here in our study, performed significantly better, especially in the virtual reality condition. We suggest that the regular manipulation of complex visual structures through virtual reality tools gives access to specific eligibility traces which enable rapid and successful many-to-one matching. Two possible interpretations of the data from our study may be suggested: 1) The eligibility traces made accessible through interaction with virtual data are specific to the expertise of architects and design engineers and involve learnt knowledge relative to Euclidian or projective geometry (specific hypotheses regarding such learnt eligibility traces are given on C. Silvestri's website at http://www.lmgc.univ-montp2.fr/~silvestri/). 2) The eligibility traces for visual-structural many-to-one matching are accessed "naturally" by anyone who regularly interacts with complex visual structures through communication with virtual reality, as in computer games, for example. Further experiments on other study populations will contribute to our understanding of the nature of the eligibility traces in perceptual-structural learning through virtual data.

**CONCLUSIONS**

Regular interaction with virtual reality data facilitates the perceptual representation of novel, complex visual structures through a perceptual learning process that makes specific eligibility traces available to enable many-to-one memory matching, as explained above. Why experts in conceptual design access these eligibility traces through brief, multiple viewing of a complex virtual structure that is new to them while non-experts do not may be explained either on the basis of their specific expertise in descriptive geometry, or the selective activation of universal geometric representations (e.g. Dehaene *et al.*, 2006, Dresp *et al.*, 2007) in anyone, expert or not, who regularly interacts with virtual environments. The latter explanation is supported by the observation that multiple viewing of the real-world model produced poorer performances within the expert population compared with multiple viewing of the virtual structure, indicating a training effect that may be specific to the privileged communication with virtual environments in that study group.

**ACKNOWLEDGEMENT:** Financial support from the I.U.A.V. in Venice to C. Silvestri is gratefully acknowledged. We are grateful to Professor Enzo Siviero for his help and advice.

Singh, S. P., & Sutton, R. S. (1996) Reinforcement learning with replacing eligibility traces. *Machine Learning*, **22**, 123-158.

Snelson, K. D., (1965) Continuous tension, discontinuous compression structures. *US Patent Number 3*, **169**, 611.

Spetch, M. L., & Friedman, A. (2003) Recognizing rotated views of objects: interpolation versus generalization by humans and pigeons. *Psychonomic Bulletin & Review*, **10**, 135-140.

Stelzer, E. M., & Wickens, C. D. (2006) Pilots strategically compensate for display enlargements in surveillance and flight control tasks. *Human Factors*, **48**, 166-181.

Whitehead, S. D., & Ballard, D. H. (1991) Learning to perceive and act by trial and error. *Machine Learning*, **7**, 45-83.